\documentclass[aps,prb,twocolumn,superscriptaddress, show pacs]{revtex4-1}
\usepackage{bm, amsmath}
\usepackage{graphicx}
\usepackage{color}
\usepackage{epstopdf}
\usepackage{graphicx}
\usepackage{dcolumn}
\usepackage{bm}
\usepackage{subfigure}

\begin{document}


\title{Stability of in-plane and out-of-plane chiral skyrmions in epitaxial MnSi(111)/Si(111) thin films:  surface twists versus easy-plane anisotropy}

\author{Andrey O. Leonov}
\thanks{A.Leonov@ifw-dresden.de}
\affiliation{Department of Chemistry, Faculty of Science, Hiroshima University Kagamiyama, Higashi Hiroshima, Hiroshima 739-8526, Japan}
\affiliation{IFW Dresden, Postfach 270016, D-01171 Dresden, Germany}

\author{Ivan M.\,Tambovtcev}
\affiliation{Department of Physics, St. Petersburg State University, St. Petersburg 198504, Russia}

\author{Igor\,S.\,Lobanov}
\affiliation{Department of Physics, St. Petersburg State University, St. Petersburg 198504, Russia}
\affiliation{Faculty of Physics and Engineering, ITMO University, 197101 Saint Petersburg, Russia}

\author{Valery\,M.\,Uzdin}
\affiliation{Department of Physics, St. Petersburg State University, St. Petersburg 198504, Russia}
\affiliation{Faculty of Physics and Engineering, ITMO University, 197101 Saint Petersburg, Russia}
\date{\today}

\begin{abstract}
{The revisited theoretical phase diagrams for thin films of cubic helimagnets with the easy-plane anisotropy are shown to have different topology as 
previously reported [Phys. Rev. B  \textbf{85}, 094429 (2012)]. 
For both in-plane and out-of-plane directions of an applied magnetic field, the phase diagrams exhibit extensive areas of stable skyrmions, which overlap for a wide range of anisotropy parameters. 
Although the existence of the out-of-plane skyrmions was contradicted within the previous theoretical models, we prove that additional surface twists lead to their stability, while the moderate easy-plane anisotropy increases the stability range of in-plane skyrmions.
Moreover, the interplay between the anisotropy and the surface twists gives rise to a stable spiral state canted with respect to the surfaces. 
Being absent in bulk helimagnets, this oblique spiral occupies vast areas at the phase diagrams in thin-film nanosystems and serves as a connecting-link between cones and helicoids.
Our theory gives clear directions for renewed experimental studies of in-plane and out-of-plane skyrmions in epitaxial MnSi(111)/Si(111) thin films. 
%
%
%

}
\end{abstract}

         
\maketitle

\vspace{5mm}

\section{Introduction}

%
Chiral magnetic skyrmions  -- nanoscale particle-like topological excitations \cite{Bogdanov94,Nagaosa,Fert} -- were first observed in MnSi, a helimagnet with the cubic chiral B20-structure  \cite{Kadowaki1982,Muehlbauer09}.  
In the bulk MnSi, skyrmions 
condense into a periodic skyrmion lattice (SkL) in a small pocket of the temperature-magnetic field phase diagram surrounded by the vast region of the conical state stability. 
It was shown that  weak interactions such as the softening of the magnetization modulus \cite{Wilhelm11,leonov2018}, dipolar interactions, fluctuations \cite{Muehlbauer09,buhrand2013} etc. grant the thermodynamical stability to the SkL in the A-phase region. 

The first direct observations of chiral skyrmions in nanolayers of cubic helimagnets (Fe$_{0.5}$Co$_{0.5}$)Si \cite{yuFeCoSi} and FeGe \cite{YuFeGe}  swerved the research focus from the bulk helimagnets to nanostructures with different geometries. 
High-symmetry nanostructured objects (like magnetic nanowires \cite{Higgins2010}, nanodisks \cite{nanodiscs,Du2015}, or nanoparticles \cite{Das2018}) provide the stabilization effect of surfaces on the skyrmion states. 
As a result, the skyrmions were observed in a  much broader range of temperatures and magnetic fields. 
Moreover, nanostructures open up the perspectives to create chiral magnetic configurations that do not exist in bulk materials, e.g., so called target-skyrmions -- skyrmions with a doubly-twisted core and a number of concentric helicoidal undulations \cite{Zheng2017,Leonov2014}.
Alongside with the low critical currents needed to set skyrmions into motion \cite{Jonietz2010,Yu2012}, the enhanced skyrmion stability in nanosystems initiated a new active research field  -- skyrmionics, which is invoked to develop a skyrmion-based magnetic memory and data processing devices \cite{Fert2013,Tomasello2014}.

\begin{figure*}
\includegraphics[width=2.0\columnwidth]{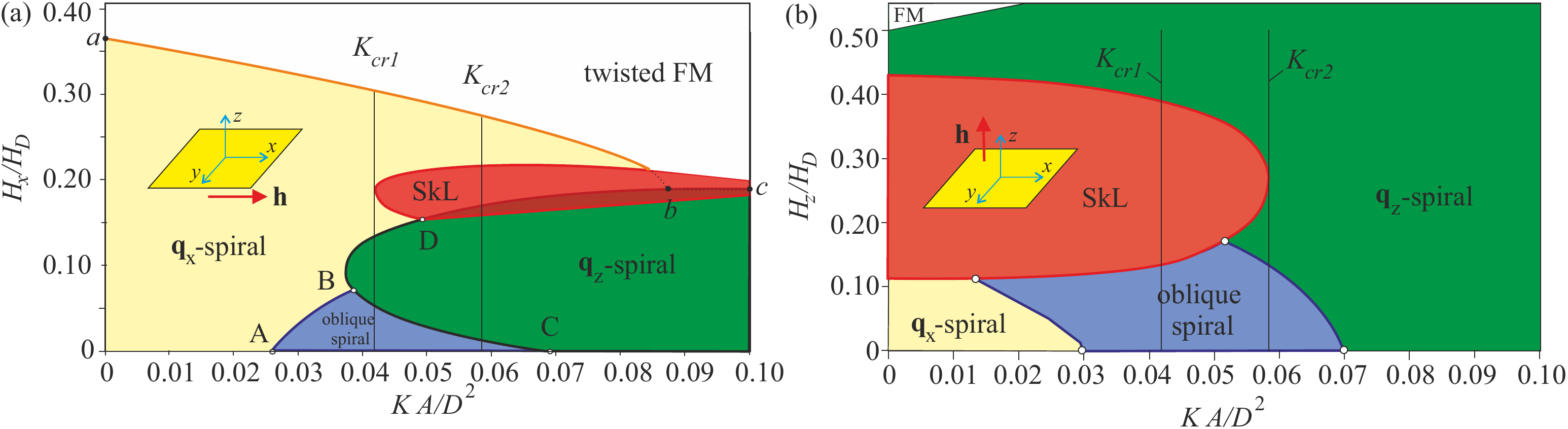}
\caption{ 
(color online) The phase diagrams of states for model (\ref{density}) plotted on the planes $(k_u,h_x)$ (a) and $(k_u,h_z)$ (b). Filled areas
indicate the regions of global stability for the SkL (red), $q_x$-spirals (yellow), $q_z$-spirals (green), and an oblique spiral (blue). White area in (a) designates the FM state with flat surface twists, whereas in (b) -- the FM state fully saturated along the field.
Solid lines stand for the phase transitions between the corresponding states. 
The critical anisotropy values indicate an overlap interval $[K_{cr1},K_{cr2}]$ with coexisting out-of-plane and in-plane skyrmions (see text for details).
\label{fig:PD}
}
\end{figure*}

Two main physical mechanisms have been proposed to date to explain the formation of skyrmion lattices in nanolayers of cubic helimagnets.
\textit{In the first mechanism} one mainly acts on the main rival state, the conical phase, trying to impair its ideal rotational configuration by small anisotropies, e.g., either by deforming it with easy and hard axes of the cubic anisotropy \cite{Leonov2019} (which is considered to be an intrinsic magnetocrystalline anisotropy) or by closing the cone with the easy axis of the uniaxial anisotropy \cite{Butenko2010} co-aligned with its wave vector (which is considered to be the surface/interface induced anisotropy). Skyrmions are more resilient to these deformations and thus gain stability in vast regions of the phase diagrams.
%
%
\textit{In the second stabilization mechanism} one modulates the internal structure of skyrmions by additional surface twists \cite{Rybakov2013}. 
Indeed in bulk helimagnets, only the Lifshitz invariants (LI) $\mathcal{L}^{(x,y)}_{x,y}$ govern  the magnetization rotation in skyrmions.
Here $\mathcal{L}^{(k)}_{i,j} = m_i \partial m_j/\partial x_k - m_j  \partial m_i/\partial x_k$ are the energy terms with the first derivatives of the magnetization $\mathbf{m}$ with respect to the spatial coordinates $x_k$.
These LIs fix the skyrmion helicity at the value $\gamma=\pi/2$ 
(Bloch-like fashion of rotation).
In magnetic nanolayers on the contrary, the LI  $\mathcal{L}^{(z)}_{x,y}$ with the magnetization derivative along $z$ comes into play. This is the energy term that stipulates the magnetization rotation within the conical phase, as well. 
For skyrmions, $\mathcal{L}^{(z)}_{x,y}$ leads to the gradual change of the skyrmion helicity ($\gamma=\pi/2\pm\delta(z)$) towards upper and lower surfaces with the penetration depth $0.1 L_D$ 
($L_D$ is the equilibrium spiral period)\cite{Rybakov2013,twists}. 
This effect accumulates additional negative energy compared with the cones not decorated by the additional surface twists \cite{Rybakov2013,twists}. Hence, SkL is stabilized in a broad range of applied out-of-plane magnetic fields and nanolayer thicknesses even without any anisotropic contributions \cite{twists}. 

The epilayers of MnSi on Si (111) substrates represent a system, in which both aforementioned stabilization mechanisms interplay \cite{Meynell2017,Meynell2014,Karhu2012,Wilson2012}. 
High-quality thin films of MnSi are grown by molecular beam epitaxy \cite{Karhu2012}. 
Owing to the lattice mismatch between the B20 crystal and the Si(111) substrate, the film is tensily strained with the strain monotonically decreasing with the film thickness.
This leads to the uniaxial anisotropy $K$ (UA) with a hard axis perpendicular to the layer. 
To stabilize skyrmions with such an easy-plane UA, one applies a magnetic field in-plane: by this, the anisotropy leads to elliptical deformation of the conical state with the wave vector along the in-plane magnetic field and thus leads to the stability of in-plane skyrmions \cite{Wilson2012} for a range of magnetic-field strengths and the values of UA larger than some threshold value, $K>K_{cr1}$ (Fig. \ref{fig:PD} (a)). 
For the out-of-plane direction of an applied magnetic field, the easy plane anisotropy, on the contrary, increases the energy of (111)-oriented skyrmions that are stable due to the surface twists and eventually suppresses them for $K>K_{cr2}$ (Fig. \ref{fig:PD} (b)). 
%
%
Thus, we note that in MnSi/Si (111) epilayers, surface effects enabling stability of out-of-plane skyrmions compete with the uniaxial anisotropy underlying stability of in-plane skyrmions.

%
%
%

The existence of the in-plane skyrmions in epitaxial MnSi/Si(111) thin films was recently unambiguously proved by the combination of polarized neutron reflectometry and small-angle neutron scattering (SANS) \cite{Wilson2012}. Such indirect experimental techniques should had been utilized since in-plane skyrmions could be hardly spotted by the conventional detection methods such as Lorentz TEM.
In Ref. \onlinecite{Yokouchi2015}, the authors also detected the characteristic planar Hall anomalies ascribed to the in-plane skyrmion strings. 
With the in-plane skyrmions being  an undisputed experimental fact, the situation with out-of-plane skyrmions in strained MnSi/Si(111) films remains rather controversial. 
In Ref. \onlinecite{Li2013} by a combination of the Lorentz TEM and measurement of the topological Hall effect, an area of out-of-plane skyrmions was revealed  over a much wider temperature-magnetic field range than the skyrmionic A-phase of bulk MnSi range \cite{Kadowaki1982,Muehlbauer09}. 
However in Ref. \onlinecite{Wilson2014}, the authors excluded any possibility for out-of-plane skyrmions in epitaxial MnSi/Si(111) thin films.
Such a radical statement was based mainly on the theoretical arguments that the particular anisotropy in MnSi/Si(111) entirely suppresses these states in an out-of-plane magnetic field, and the cone phase is the only stable magnetic texture below the saturation field. The corresponding theoretical phase diagram, although for a bulk MnSi with the easy-plane UA, was constructed in Fig. 3 of Ref. \onlinecite{Wilson2014}. 
Moreover, the authors underpinned their theoretical arguments by the experimental results: they discovered no first-order magnetic phase transitions that would signal the appearance of (111) skyrmions \cite{Wilson2014}.

In this manuscript, we revisit the theoretical phase diagrams for thin films of cubic helimagnets with the easy-plane UA and both directions of an applied magnetic field.
We argue that the topology of the phase diagrams is different from that constructed for a bulk MnSi in Ref. \onlinecite{Wilson2014}.
We show that both types of skyrmion states are realized in a wide range of anisotropy values with the significant overlap $[K_{cr1},K_{cr2}]$ (Fig. \ref{fig:PD}). We focus on the internal properties of the SkL states as well as on their field-driven evolution and instabilities.
We also report a novel oblique spiral state that occupies an extensive area of the phase diagrams for nanolayers, but does not exist for bulk  cubic helimagnets. We determine the equilibrium parameters of the oblique spiral as functions of applied fields and/or the values of UA and speculate that this state can be considered as an intermediate state between two conventional spirals with the wave vectors parallel and perpendicular to the film.
We identify first-order phase transition between the oblique spiral and its neighbors on the phase diagrams, which has not been indicated so far by the experiments. 


%
\begin{figure*}
\includegraphics[width=1.99\columnwidth]{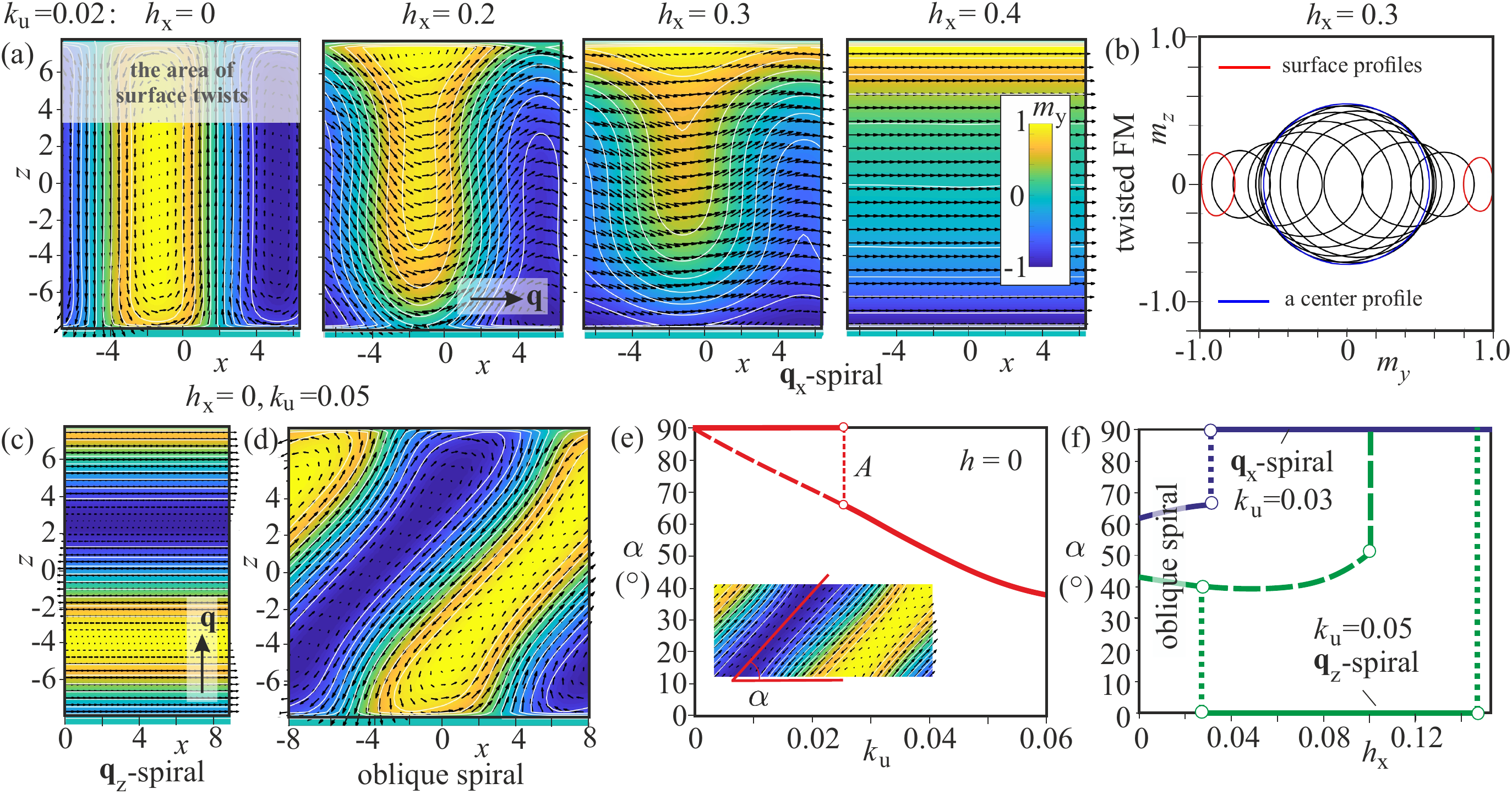}
\caption{ 
(color online) 1D spiral states in an in-plane magnetic field. The color plots stand for $m_y$-components of the magnetization. 
$m_x$- and $m_z$-components are shown with thin black arrows. (a) A field-driven transformation of a $q_x$-spiral into a twisted FM state. (b) A set of $m_y-m_z$ traces through the film thickness: 
blue line designates a profile in the layer middle ($z=0$), red curves -- at both surfaces ($z=\pm L/2$). (c) A $q_z$-spiral has just one loop of the magnetization rotation, which is unwound by transforming into the twisted FM state (line $b-c$ in Fig. \ref{fig:PD} (a)) or into a $q_x$-spiral (line $B-D-b$ at the phase diagram). For $h_x=0$ (at the point $A$ (e)), a $q_x$-spiral gives rise to an oblique spiral (d) with some tilt angle $\alpha$ (inset of (e)) that smoothly decreases depending on $k_u$ (e) up to its transition into a $q_z$-spiral. (f) Field-driven evolution of the oblique spiral for $k_u=0.03$ (blue curves) and $k_u=0.05$ (green curves) exhibits its jumps into conventional conical and helicoidal spirals (see text for details).
\label{fig:oblique}
}
\end{figure*}

\section{Phenomenological model}

The standard model for magnetic states in cubic non-centrosymmetric ferromagnets is based on the Bak-Jensen functional \cite{Dz64,Bak80} that includes an additional uniaxial anisotropy with constant K:
\begin{equation}
w =A\,(\mathbf{grad}\,\mathbf{m})^2 + D\,\mathbf{m}\cdot \mathrm{rot}\,\mathbf{m} -\mu_0 \,M  \mathbf{m} \cdot \mathbf{H} + Km_z^2,
\label{density}
\end{equation}
The principal interactions essential  to stabilize modulated states are as follows:  the exchange stiffness with constant $A$,  Dzyaloshinskii-Moriya interaction (DMI)  with constant $D$, and the Zeeman energy. 
$\mathbf{m}= (\sin\theta\cos\psi;\sin\theta\sin\psi;\cos\theta)$ is the unity vector along the magnetization vector  $\mathbf{M} = \mathbf{m} M$, and $\mathbf{H}$ is the applied magnetic field.
We investigate Eq. (1) for both orientations of the applied magnetic field, out-of-plane $\mathbf{H}||z$ and in-plane  $\mathbf{H}\perp z$.
$K > 0 $, which stands for an easy-plane anisotropy or equivalently for a hard axis perpendicular to the film.

The equilibrium magnetic states 
are derived by the Euler equations for the energy functional (\ref{density}) together with the Maxwell equations and with corresponding boundary conditions. The solutions depend on the two control parameters of the model (\ref{density}), the  reduced 
 magnetic field,  $h = H/H_D$ and the value of the UA, $k_u=KA/D^2$. Here, $L_D = A/|D|$  is the characteristic length unit of the modulated states. In the following, the spatial coordinates are measured in units of $L_D$. The value $4\pi L_D$ for $H=K=0$ is the \textit{helix period} for bulk helimagnets ($18nm$ for the bulk MnSi).   
$\mu_0 H_D = D^2/(A M)$ is the \textit{critical field}. For a conical phase in bulk helimagnets, the saturation field in units of $H_D$ equals $h = 0.5$. 

In this paper, we neglect effects imposed by spatial inhomogeneity of the induced anisotropy. 
%
The considered film is infinite in $x-$ and $y-$ directions, i.e., we apply the periodic boundary conditions.
The thickness of the film is set to $L =1.2 L_D$ to be comparable just with one row of in-plane skyrmions,  and the film is confined by the parallel planes at $z=\pm 0.6 L_D$. 
The cubic anisotropy and the anisotropic exchange are omitted in functional (\ref{density}) although recently they have been found to stabilize SkL even in bulk helimagnets \cite{Chacon2018,Bannenberg2019}.

In the following discussion to avoid any ambiguity, we will refer to $q_x$- and $q_z$-spirals with the corresponding orientation of their wave vectors. In the introduction, however, we adhered to the terminology related to the direction of the field, i.e., we called $q_z$-spiral a conical state for $\mathbf{H}||z$ ($q_x$-spiral being a helicoid in this case), whereas for $\mathbf{H}\perp z$ the $q_x$-spirals were called cones ($q_z$-spiral being a helicoid).

\begin{figure*}
\includegraphics[width=2.0\columnwidth]{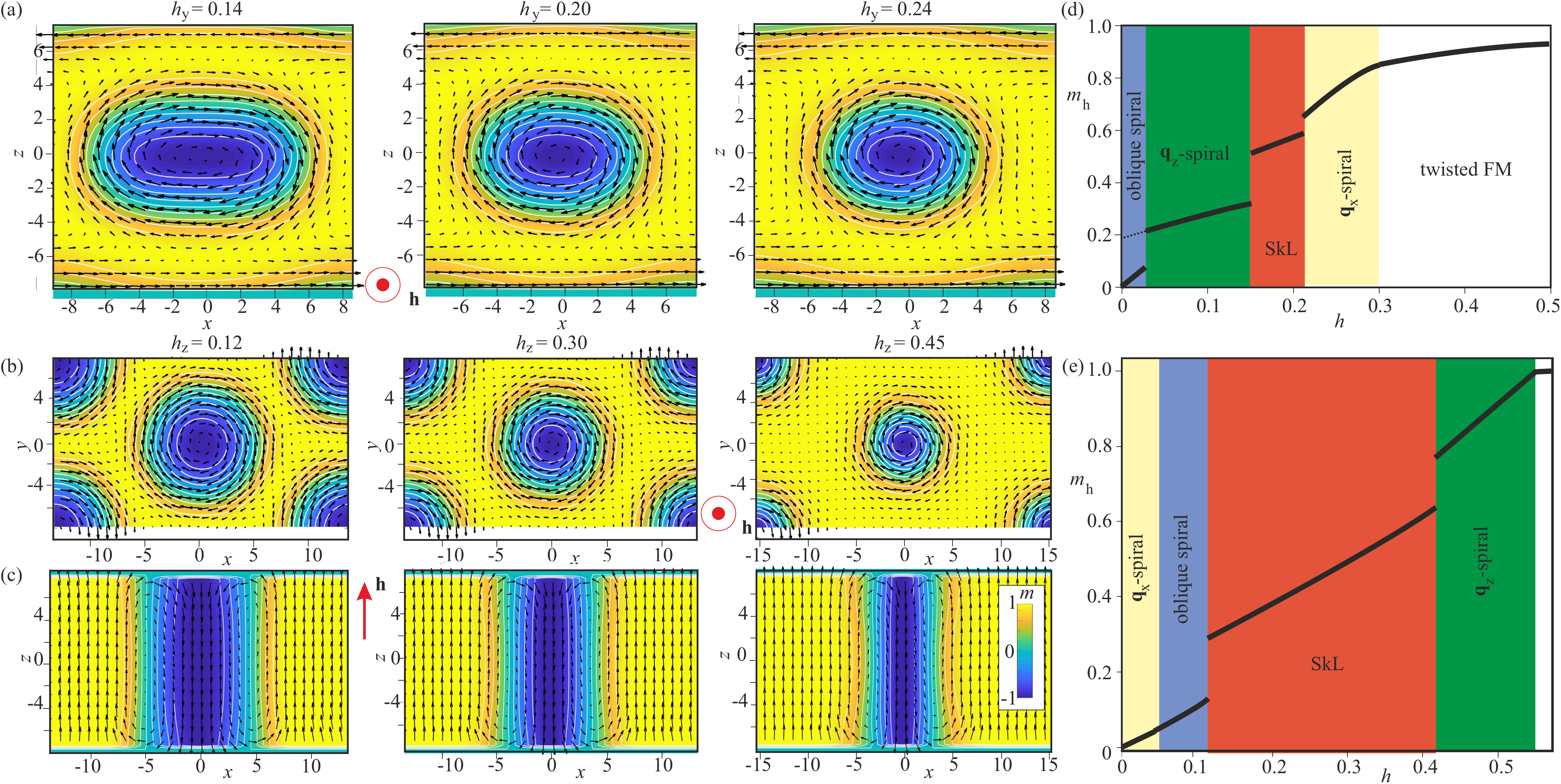}
\caption{ (color online) Magnetic structure of in-plane skyrmions, $k_u=0.05$ (a),  showing their elliptical instability in an applied magnetic field and out-of-plane skyrmions, $k_u=0.025$ (b), (c), showing the conventional SkL period expansion in an increasing field. 
The internal structure of out-of-plane skyrmions is shown within two mutually perpendicular planes $xy$ (b) and $xz$ (c).
Magnetization curves $m-h$  for the in-plane (d) and the out-of-plane (e) orientations of the field demonstrate jumps at the lines of the phase transitions according to the phase diagrams in Fig. \ref{fig:PD}.
\label{fig:skyrmions}
}
\end{figure*}

\section{The phase diagrams of states}

%
By comparing the equilibrium energies of 1D spiral states, 2D SkL and polarized FM states, we construct the phase diagrams (PD) on the planes $(k_u,h_x)$ (Fig. \ref{fig:PD} (a)) and $(k_u,h_z)$ (Fig. \ref{fig:PD} (b)). 
The regions of modulated states corresponding to the global minimum of the energy functional (\ref{density}) are indicated by different colors. Note, that the FM state magnetized along the in-plane magnetic field (white area in Fig. \ref{fig:PD} (a)) acquires the flat surface twists (the color plot for $h_x=0.4$ in Fig. \ref{fig:oblique} (a)) that never fully saturate \cite{Meynell2014}. For the out-of-plane magnetic field, the FM state is fully saturated \cite{Butenko2010} (white area in Fig. \ref{fig:PD}  (b)) and is a result of the cone closing at the critical field, $h_{z0}=(1+4k_u)/2$. 

\subsection{$q_x$-spirals}
%
In Fig. \ref{fig:PD} (a), the $q_x$-spirals occupy the yellow-shaded area of the PD. For $h_x = k_u = 0$ due to the additional surface twists in thin-film nanosystems (indicated in Fig. \ref{fig:oblique} (a)), such spirals have lower energy as compared with the $q_z$-spirals. 
Thus the region $0-A$ needed for the easy-plane UA to outbalance the energy of the spiral surface twists and to favor other 1D states is a noticeable feature of the phase diagrams. 
For bulk helimagnets, the point $A$ 
tends to zero as constructed in Fig. 2 of Ref. \onlinecite{Karhu2012}. 
The field-driven transformation of the $q_x$-spiral  into the twisted FM state at the line $a-b$ (Fig. \ref{fig:PD} (a)) is shown in Fig. \ref{fig:oblique} (a) for an in-plane magnetic field co-aligned with the $\mathbf{q}$-vector  (see Appendix on the spiral transformation in the field $h_y$).
Besides such a wiggling structure in the plane $xz$  connecting the cone-like fashion of rotation in the middle of the film and the surface twists, the uniaxial anisotropy causes slight elliptical distortions of the $q_x$-spirals in the $yz$ plane. Fig. \ref{fig:oblique} (b) exhibits the traces of the magnetization rotation by moving from the middle of the layer (blue curve) to the confining surfaces (red curves).

\subsection{$q_z$-spirals}

%
$q_z$-spirals in this geometry are oriented perpendicular to the field and inhabit 
the green-shaded region at the phase diagram (Fig. \ref{fig:PD} (a)).  Since usually the film thickness is a non-integer multiple of the helical wavelength (1.2 in the present paper), the magnetization of the $q_z$-spiral has an  uncompensated value even in zero magnetic field (see dotted line as a continuation of the magnetization curve for a $q_z$-spiral in Fig. \ref{fig:skyrmions} (b)).
The solutions for this field-distorted spiral are obtained from the well-known differential equations for the non-linear pendulum \cite{Dz64,Wilson2012}.  Experimentally, one indicates a set of first-order phase transitions related to the unwinding processes of  spiral loops in an applied  magnetic field \cite{Meynell2014}. The line $b-c$ indicates a corresponding transition between the $q_z$-spiral (Fig. \ref{fig:oblique} (c)) and the twisted  FM state. 
Otherwise, the green area of the phase diagram is free from this sort of phase transitions.
Along the transition line $B-D-b$, however, the twisted FM state should acquire additional undulations and transform into a $q_x$-spiral  shown in Fig. \ref{fig:oblique} (a) (contour plots for $h_x=0.2, 0.3$).
At the line $B-C$ by the first-order phase transition, $q_z$-spiral acquires a tilt. 

\subsection{Oblique spirals}

%
The thin-film geometry also permits a stable oblique spiral (Fig. \ref{fig:oblique} (d)), which originates from the interplay between the surface twists and the easy-plane anisotropy \cite{Tambovtcev}: whereas the negative energy associated with the surface twists remains almost unchanged, the canting leads to the lowering of the positive anisotropy energy (for the details on the calculations see Appendix).
The angle of canting $\alpha$ for $h_x=0$ (inset of Fig. \ref{fig:oblique} (e)) 
monotonically decreases with the growing UA. However up to the point $A$ (dashed line), the oblique spiral is a metastable solution as compared with the "straight" spiral (solid line). At $k_u=0.069$, the oblique spiral  undergoes a transition into the $q_z$-spiral with $\alpha=0$.

Fig. \ref{fig:oblique} (f) shows the field evolution of the canting angle  for $k_u=0.03$ (blue curves) and $k_u=0.05$ (green curves), which reflect abrupt jumps between the corresponding stable spiral states (solid lines).
Moreover for $k_u=0.05$ the field driven evolution may occur along two different paths. The first option is shown by the dotted and solid green lines. The oblique spiral jumps into the $q_z$-spiral with $\alpha=0$, which eventually jumps into the $q_x$-spiral with $\alpha=90^{\circ}$. All the transitions occur at the boundaries of the corresponding stability regions at the phase diagram (Fig. \ref{fig:PD} (a)). In the second option (shown by the dashed green lines), however, the oblique spiral may last out as a metastable state with the higher energy and then jump directly into the $q_x$-spiral by omitting the region with the $q_z$-spiral. 

For both field directions, the stability region of this spiral is quite extensive and settles in the direct vicinity of other spiral states. This can be the reason that experimentally such a state has not been identified yet: 
the first-order phase transition of this spiral into the conical state is easy to misinterpret as a magnetization process related to the $q_z$-spiral.

\section{Skyrmion tubes}


\subsection{In-plane skyrmions}

%
Skyrmions for the in-plane direction of the field are sandwiched between flat surface twists at both surfaces and are confined to the middle of the film due to the potential well formed by the surface twists.
In such a geometry, the surface twists mainly do not impact the internal structure of skyrmions.
Due to the UA with its easy-plane $xy$ dissecting skyrmion tubes parallel to the film surfaces, the skyrmions  undergo an elliptical instability (Fig. \ref{fig:skyrmions} (a)). This is also the reason that skyrmions, besides the region of their thermodynamical stability (red-shaded area in Fig. \ref{fig:PD} (a)), are sustainable only in a very narrow parameter range. 
In general, skyrmions in multiple rows occupy the cross-section of the film \cite{Keesman2015}. 
Then upon tuning the magnetic field through the skyrmion phase, the system would exhibit a cascade of first-order phase transitions through the states with the different number of skyrmion rows \cite{Keesman2015}. 

\subsection{Out-of-plane skyrmions}

%
As mentioned in the introduction,  the solutions for skyrmions and $q_x$-spirals  are modulated along the film thickness for the out-of-plane magnetic field \cite{Rybakov2013,twists}. 
In some sense, such solutions can be considered as a superposition of corresponding solutions in bulk helimagnets and specific twisted modulations near the surfaces that involve all rotational terms $\mathcal{L}^{(k)}_{i,j}$.

For $k_u = 0$, the following first-order phase transitions are identified at the theoretical phase diagrams for thin layers of cubic helimagnets: $q_x$-spiral--SkL, SkL--$q_z$-spiral, $q_z$-spiral--FM state \cite{twists,Rybakov2016}. Here, we omit the regions of the phase diagrams related to different surface states as chiral bobbers and/or stacked spirals \cite{Rybakov2016}.
For $k_u=0.02$, two additional transitions are present: the oblique spiral--SkL and $q_x$-spiral -- oblique spiral. 
The latter transition, however, is hardly identified at the magnetization curves (Fig. \ref{fig:skyrmions} (d)).

In free-standing layers, SkLs stabilized by the mechanism of surface twists  were predicted to exist up to very large film thicknesses ($L/4\pi L_D\approx 8$) \cite{twists,Rybakov2016}.
In FeGe free-standing wedges, however, SkL was observed experimentally  only when the thickness is smaller than $\approx 130 nm$, which is less than 2 in the units of $4\pi L_D$. 
The reduced effect of surface twists was explained by the temperature dependence of the material parameters. 
In epitaxial MnSi(111)/Si(111), on the contrary, a good agreement was found between the numerical simulations for the flat surface twists and their experimental realization \cite{Meynell2014}. 
Thus, the influence of the surface modulations on the out-of-plane skyrmions might also turn out to be unimpeded and lead to the skyrmion stability \cite{twists,Rybakov2016}.
An overlap interval $[K_{cr1},K_{cr2}]$ also implies that skyrmions can be found for canted magnetic fields. This would also constitute an experimental strategy to observe the out-of-plane skyrmions: for the fixed value of an in-plane magnetic field in the region of skyrmion stability, one would rotate the field to transit into the region of out-of-plane skyrmions for the same value of the field. 
We also do not exclude formation of  skyrmion clusters with mutually orthogonal skyrmion tubes  recently reported in chiral liquid crystals \cite{Sohn2019,Vlasov2020}. 
%
%

\section{Conclusion}

%
In conclusion, robust and thermodynamically stable skyrmions can be induced for the out-of-plane orientation of the magnetic field. 
To suppress the effect of the chiral surface twists, relatively large values of the easy-plane anisotropy are needed ($k_u>0.058$, i.e. larger than the critical value $K_{cr2}$) as compared in particular with the anisotropy needed for the stability of elliptically distorted in-plane skyrmions ($k_u>0.042$, i.e. larger than the critical value $K_{cr1}$). Thus, both types of SkLs co-exist in some anisotropy parameter range $[K_{cr1},K_{cr2}]=[0.042,0.058]$.
However, additional analysis is required for the evolution of the phase diagrams with the layer thickness. Indeed, the impact of the surface twists decreases with the growing thickness what would mean diminishing area of skyrmion stability for the out-of-plane magnetic fields and the lower value of $K_{cr2}$. 
At the same time, the value of $k_u$ due to the tensile strain also decreases what makes out-of-plane skyrmions feasible up to relatively large thicknesses. 
The influence of demagnetizing effects that were neglected in the present paper should also be studied systematically especially for the out-of-plane direction of the field. 

Moreover, we found a new low-field spiral state that intervenes between the conventional conical and helicoidal states and is stable in a broad parameter range.
Such an oblique spiral constitutes an interesting deviation from the previously published phase diagrams for bulk helimagnets, which may have important consequences for the field of chiral magnetism. In particular, the spiral tilt can also give rise to new topological magnetic defects, such as isolated skyrmions, with interesting static and dynamic properties.

In total, our findings underscore the paramount role of magnetic anisotropies and surface twists in stabilizing skyrmionic states with different orientation and provide valuable directions to manipulate and tune skyrmions in real experiments, e.g., in epitaxial MnSi(111)/Si(111) thin films. 
Recent experiments on strained itinerant helimagnet FeGe \cite{Ukleev2020} also demonstrate a promising role of tensile strain for the creation and manipulation of magnetic solitonic textures: the theoretical concepts of the present manuscript could also be tested in this system.

\section{Acknowledgements. }
%
This work was funded by Russian Science Foundation (Grant 19-42-06302). 
AOL thanks Ulrike Nitzsche for technical assistance.

\section{Appendix: solutions for an oblique spiral state}

\subsection{Energy minimization procedure for an oblique spiral}

\begin{figure*}
\includegraphics[width=2.0\columnwidth]{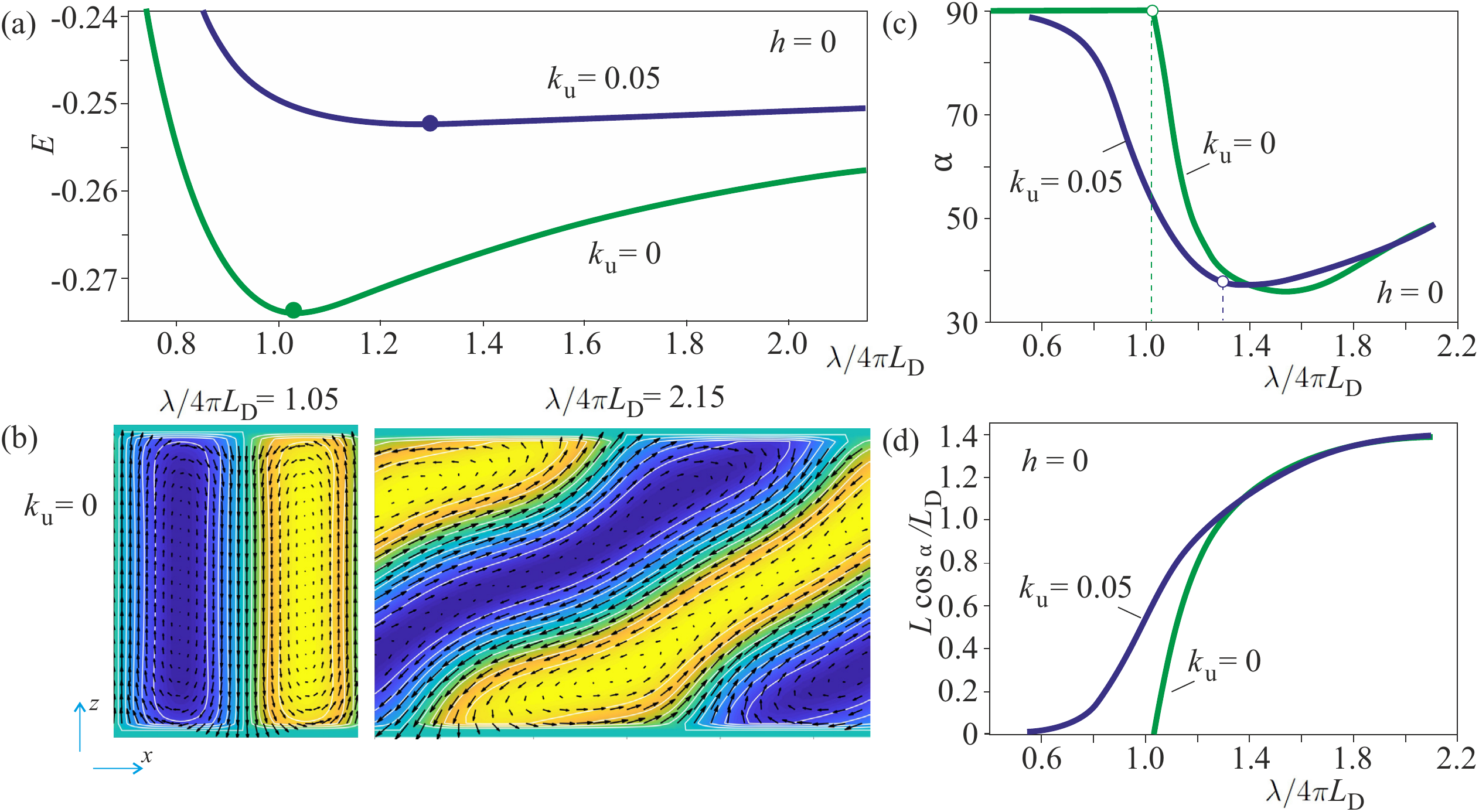}
\caption{ 
(color online) Energy minimization procedure for an oblique spiral (see text for details).
\label{fig:minimization}
}
\end{figure*}

The energy density of an oblique spiral is minimized with respect to the spiral period along $x$. For each point $(k_u,h)$ at the phase diagrams (Fig. 1) we find an equilibrium spiral period. 
Fig. \ref{fig:minimization} (a) shows the zero-field energy densities for $k_u=0$ (green curve) and $k_u=0.05$ (blue curve) depending on the spiral period. The canting angle of the oblique spiral varies along these energy curves. 
Interestingly for $k_u=0$, the spiral remains "straight" (Fig. \ref{fig:minimization} (b)) in the range $\lambda/4\pi<1.05 L_D$, i.e. up to the energy minimum shown by the green circle. However, at $\lambda/4\pi>1.05 L_D$, a spiral becomes oblique (Fig. \ref{fig:minimization} (b)) with the canting angle growing with the spiral period. By this, the oblique spiral tries to retain its equilibrium period determined by the exchange and DMI. 
This can be judged also by the projection of the spiral period $\lambda \cos \alpha/ /4\pi L_D$ (Fig. \ref{fig:minimization} (d)).
For $k_u=0.05$ the equilibrium period is achieved at a slightly higher value 
(blue curve in Fig. \ref{fig:minimization} (a)),
and the canting angle varies along the whole energy curve 
(Fig. \ref{fig:minimization} (c)).
%
%

\begin{figure*}
\includegraphics[width=2.0\columnwidth]{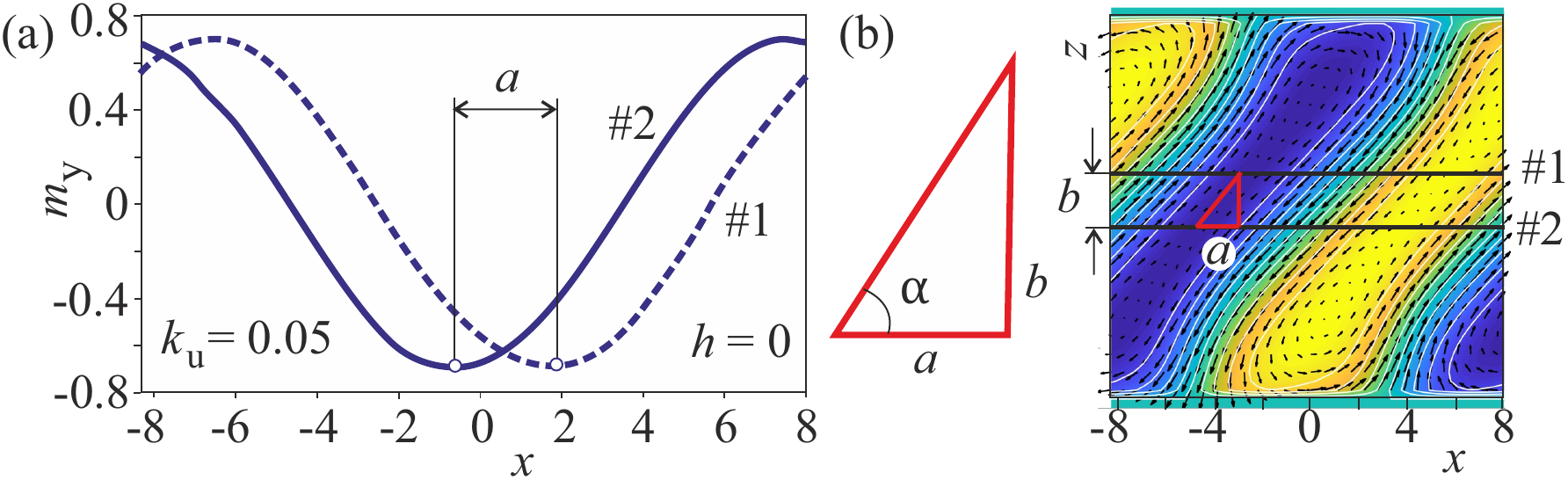}
\caption{ 
(color online) A procedure to determine a canting angle for an oblique spiral state (see text for details).
\label{fig:angle}
}
\end{figure*}


\subsection{A procedure to determine a canting angle for an oblique spiral state}

An oblique spiral represents a combination of a flat spiral with some canting angle $\alpha$ in the middle of the layer and a roundish part near the surfaces related to the chiral surface twists (Fig. \ref{fig:angle} (b)). By varying its canting angle, the oblique spiral optimizes  
the impact of the surface twists and an easy-plane anisotropy.
To introduce a procedure for defining a canting angle, we consider two profiles of the $m_y$-components of the magnetization located at some fixed distance $b$ from each other near the layer middle (Fig. \ref{fig:minimization} (a)).
In this undistorted part of an oblique spiral, these profiles are essentially the same but acquire a phase shift $a$ with respect to each other. 
Thus, the canting angle is defined as $\tan \alpha=b/a$ (Fig. \ref{fig:minimization}, inset). 
For a straight $q_x$-spiral $\alpha=90^{\circ}$, for a $q_z$-spiral -- $\alpha=0$.

\begin{figure*}
\includegraphics[width=1.5\columnwidth]{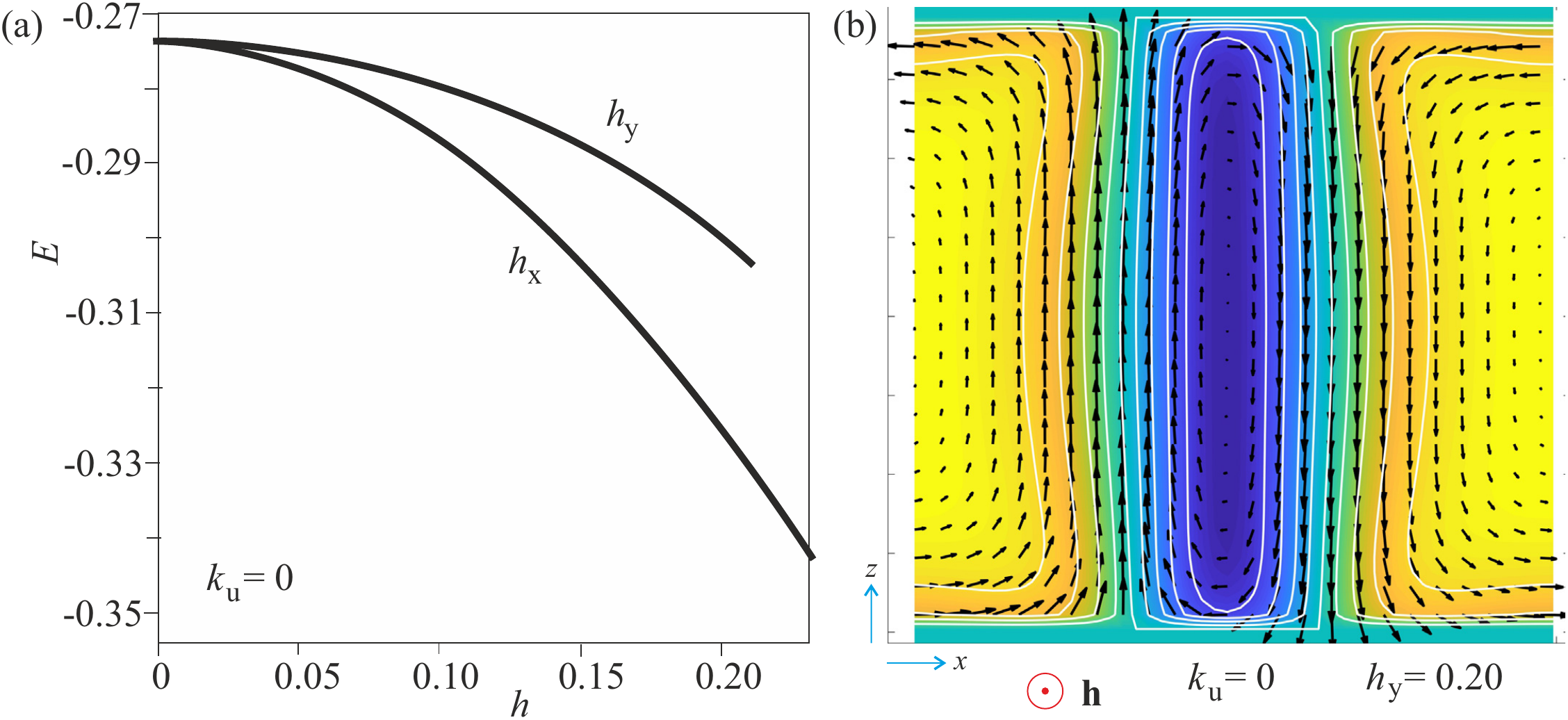}
\caption{ 
(color online) A field-driven evolution of $q_x$-spiral for $\mathbf{H}||y$ (see text for details).
\label{fig:hy}
}
\end{figure*}


\subsection{A field-driven evolution of $q_x$-spiral for $\mathbf{H}||y$}

In the main part of the manuscript, we considered 1D spiral states only for an applied magnetic field coaligned with the in-plane $\mathbf{q}$-vectors, i.e., $\mathbf{H}||x$.
The energy of such states has been found to be lower as compared with the corresponding states for the field perpendicular to their in-plane wave vectors. 
As an example, Fig. \ref{fig:hy} (a) shows the energy densities for $q_x$-spirals and both direction of the field. 
The $q_x$-spiral gradually expands (Fig. \ref{fig:hy} (b)) in the field $h_y$ as would also be the case in bulk helimagnets.

\begin{figure*}
\includegraphics[width=1.5\columnwidth]{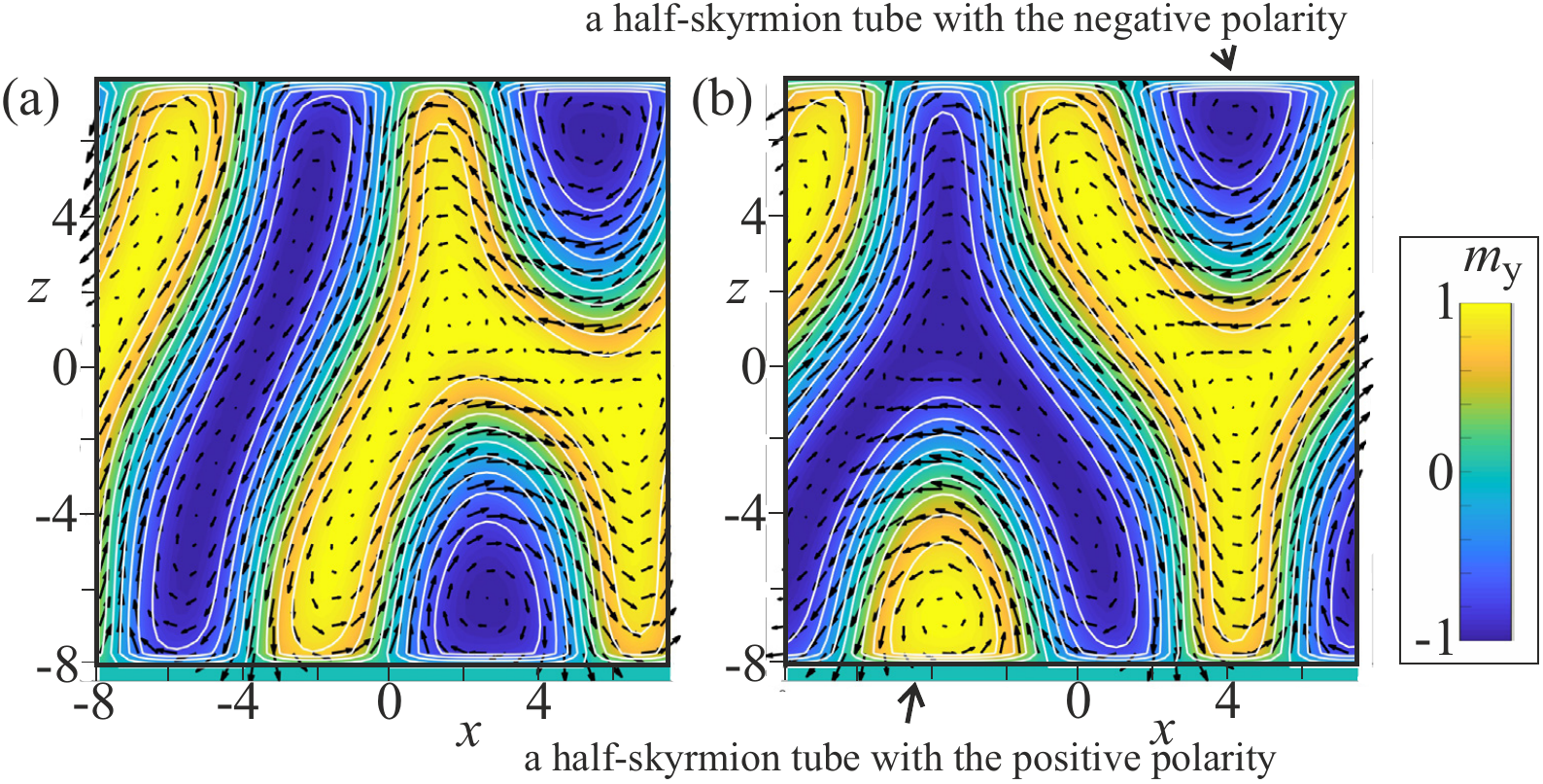}
\caption{ 
(color online) Oblique spiral states with two senses of canting (see text for details).
\label{fig:DW}
}
\end{figure*}


\subsection{Oblique spiral states with two senses of canting}

An oblique spiral may cant equivalently along both directions. Then, one could envision some sort of Y-shaped domain walls between domains of an oblique spiral with a particular sense of canting (Fig. \ref{fig:DW}). Such a shape enables formation of half skyrmions running parallel to the surface and having both polarities.

\end{document}